\documentclass[twocolumn,bibtex]{biophys-new}
\usepackage[utf8]{inputenc}
\usepackage{graphicx}
\usepackage[colorlinks,allcolors=cyan!70!black]{hyperref}

\newcommand{\front}{\mathrm{f}}
\newcommand{\back}{\mathrm{b}}
\newcommand{\taumig}{\tau}
\newcommand{\rbfi}{\mathbf{r}_i^{\back \front}}

\newcommand{\Ncell}{N}
\newcommand{\Rmax}{R_{\mathrm{max}}}
\newcommand{\DT}{\mathcal T}

\newcommand{\Fmot}{\mathbf{F}_{\mathrm{mot}}}

\title{Spontaneous Spatiotemporal Ordering of Shape Oscillations Enhances Cell Migration}
\runningtitle{Shape Oscillations Enhance Migration} 

\author[]{M. Campo}
\author[]{S.K. Schnyder}
\author[]{J.J. Molina}
\author[*]{T. Speck}
\author[]{R. Yamamoto}
\runningauthor{Campo et al}

\corrauthor[*]{thomas.speck@uni-mainz.de}

\papertype{Article}

\begin{document}

\begin{frontmatter}

\begin{abstract}
The migration of cells is relevant for processes such as morphogenesis, wound healing, and invasion of cancer cells. In order to move, single cells deform cyclically. However, it is not understood how these shape oscillations influence collective properties. Here we demonstrate, using numerical simulations, that the interplay of directed motion, shape oscillations, and excluded volume enables cells to locally ``synchronize'' their motion and thus enhance collective migration. Our model captures elongation and contraction of crawling ameboid cells controlled by an internal clock with a fixed period, mimicking the internal cycle of biological cells. We show that shape oscillations are crucial for local rearrangements that induce ordering of internal clocks between neighboring cells even in the absence of signaling and regularization. Our findings reveal a novel, purely physical mechanism through which the internal dynamics of cells influences their collective behavior, which is distinct from well known mechanisms like chemotaxis, cell division, and cell-cell adhesion.
\end{abstract}

\end{frontmatter}

\section*{Introduction}

The collective migration of cells plays an essential role for several biological processes including the formation of embryos, the closure of wounds, and metastasis of tumor cells~\citep{friedl2009collective,rorth2009collective,scarpa2016collective}. Motile cells convert available free energy into directed motion, and can thus be regarded as a type of active matter~\citep{ramaswamy2010mechanics, needleman2017active}. Unlike equilibrium passive systems, active matter can exhibit collective dynamic behavior such as swarming and clustering~\citep{vicsek2012collective,cates2015motility}, the study of which has gained much attention from physicists over the last decade. The essence of active systems can be captured with rather minimalistic models, such as the Vicsek model~\citep{vicsek1995novel} in which isotropic particles interact by aligning their velocity with the average velocity of their neighbors.

Biological cells possess features that are not captured by simple active particles. In fact, cells have an internal structure and a variable shape, they can have different internal states, they can interact through long-range chemical signals, and they can change their phenotype in response to external perturbations.
In addition, cells in a multicellular organism differentiate into distinct types during development, leading to different properties corresponding to their biological function. Nevertheless, motile cells can be broadly classified into two types according to their migratory phenotype. Mesenchymal cells migrate slowly and protrude their body into multiple competing lamellipodia~\citep{bear2014directed}. Amoeboid cells migrate quickly, cyclically extending their body into a well polarized shape, and have a larger persistence of motion~\citep{friedl2001amoeboid,d2017contact}. This distinction is not intrinsic as some cells are able to transition between these two migration types, such as tumor cells during invasion~\citep{pavnkova2010molecular, talkenberger2017amoeboid}.

This biological diversity makes it challenging to develop models that are able to reproduce certain phenomena of interest, while remaining general and minimalistic. Several physics-based representations of cells have been proposed in the literature~\citep{camley2017physical,hakim2017collective}. Most prominently, deformable active particles~\citep{ohta2009deformable,menzel2012soft,tarama2013dynamics,ohta2017dynamics,tjhung2017discontinuous}, vertex~\citep{farhadifar2007influence,bi2015density} or Voronoi models~\citep{honda1978description,bi2016motility}, phase field models~\citep{shao2010computational,lober2014modeling, ziebert2016computational, molina2018mechanosensitivity}, active gel models~\citep{callan13,prost15}, and subcellular element models~\citep{sandersius2008modeling,basan2013alignment,zimmermann2016contact,schnyder2017collective,tarama2018mechanics} have been investigated. Among these, subcellular element models offer the advantage of naturally taking into account the internal structure and internal forces, as well as shape deformations within a single cell. This opens up the possibility for studying the effects of such features on large aggregates of cells. To date, studies on collective cell migration have focused primarily on the roles of long-range chemical signaling~\citep{weijer2004dictyostelium,aman2008wnt}, shape~\citep{ohta2009deformable,menzel2012soft,tarama2013dynamics,ohta2017dynamics,bi2015density,manning2010coaction}, cell-cell~\citep{bershadsky2003adhesion,du2005force,trepat2009physical,manning2010coaction,bi2015density,tambe2011collective} and cell-substrate adhesions~\citep{angelini2010cell}, or cell proliferation~\citep{ranft2010fluidization,puliafito2012collective}. To our knowledge, no investigation has addressed directly the role of cyclic shape oscillations on the coherent migration of cells.

We start from the subcellular element model devised in~\citep{schnyder2017collective}, and we focus on modeling the class of motile cells that exhibit ameboid movement. The model features a two-particle representation for each cell, with an active force proportional to its length. Such an active force is able to model Contact Inhibition of Locomotion (CIL), a complex molecular process that was found to play a key role in collective migration of cells~\citep{abercrombie1953observations,mayor2010keeping}, for which protrusions are retracted upon exchanging chemical signals during cell-cell contact. Here we explicitly model the crawling dynamics, in which the active force and a contracting force alternate in order to mimic the cycle of expansion and contraction of ameboid cells.
Thanks to this explicit crawling,  we unveil a general mechanism by which conformational shape changes in ameboid cells give rise to a purely physical mechanism that enhances the collective migration of the whole tissue.


\section*{Methods}
\subsection*{Model}
\begin{figure}[t!]
  \centering
  \includegraphics[width=0.4\textwidth]{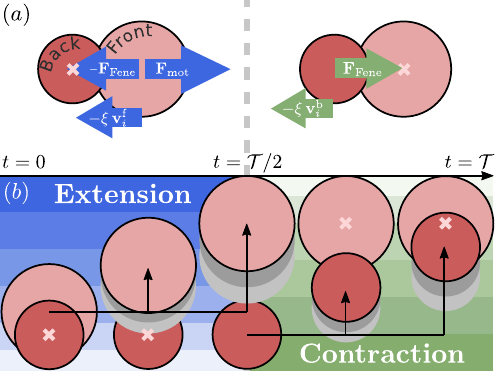}
  \caption{(a) Forces acting on a cell during expansion (left) and contraction (right), with $\Fmot$ the extensional force and $\mathbf{F}_{\mathrm{Fene}}$ the intracellular force, see \emph{Interactions} for details. (b) Illustration of the cyclic crawling: during extension the active force $\Fmot$ drives the front disk forward while the position of the back disk is fixed; during contraction the front disk is fixed while the internal force $\mathbf{F}_{\mathrm{Fene}}$ contracts the back disk towards the front. See also Supplementary Fig.~1 and Supplementary Movie 1.}
  \label{fig:illustration}
\end{figure}

Cells are represented by two disks, referred to as \emph{front} and \emph{back}, see Fig.~\ref{fig:illustration}.
The dynamics of each disk is modeled to be overdamped, which is a good approximation for a large class of cells moving in a viscous environment and dominated by active forces~\citep{tanimoto2014simple}.
The equations of motion for the front and back disks of cell $i$ read
\begin{equation}
\begin{cases}
  \xi \mathbf{v}_i^{\mathrm{b}} 
 &= \Fmot (\mathbf{r}_i; s_i)
    + \mathbf{F}_i^{\mathrm{b}},\\
  \xi \mathbf{v}_i^{\mathrm{f}} 
  & = -\Fmot(\mathbf{r}_i; s_i)
  + \mathbf{F}_i^{\mathrm{f}},
\end{cases}       
\label{eq:eq_motion}
\end{equation}
where the forces on the right hand side are decomposed into passive forces $\mathbf{F}_i^\alpha$ (for details see \emph{Interactions}) and the cell extensional force dipole $\Fmot$.
Here, $\mathbf{v}_i^{\alpha}$ is the velocity vector, $\xi$ is the friction coefficient, and $\mathbf{r}_i = \mathbf{r}^{\mathrm{f}}_i - \mathbf{r}^{\mathrm{b}}_i$ is the vector connecting the back disk to the front disk. Finally, $s_i$ indicates the stage of cell $i$, which can be either \emph{extension} ($s_i = 0$) or \emph{contraction} ($s_i = 1$). During the former, the extensional dipole force $\Fmot$ is switched on and the position of the back disk is kept fixed
\begin{equation}
  \mathbf{v}_i^{\mathrm{b}} = \mathbf{0}, \qquad
  \Fmot(\mathbf{r}; 0)
  = m\mathbf{r}
\label{eq:extension}
\end{equation}
with $m$ being a constant parameter called \emph{motility} that we fix to unity. The active extensional dipole force in Eq. (\ref{eq:extension}) drives the front disk forward along the direction of the cell's axis, so that the cell can effectively extend its length. During contraction, $\Fmot$ is switched off and the front disk is kept fixed
\begin{equation}
  \mathbf{v}_i^{\mathrm{f} } =  \mathbf{0}, \qquad
  \Fmot(\mathbf{r}; 1) = \mathbf{0}
  \label{eq:contraction}
\end{equation}
so that the internal forces within the cell move the back disk towards the front, resulting in a contraction of the cell.
The extension and the contraction stages are performed cyclically over a period $\DT$. At the beginning of the simulation each cell $i$ is randomly assigned a starting stage $s_i \in \{0, 1\}$ and a time $t_i \in \{ 0, \DT / 2 - 1 \}$ tracking the progression of the cell within its current stage.
Such cyclic dynamics induces a crawling motion [Fig.~\ref{fig:illustration}], aimed at mimicking the more complex motion of ameboid cells which is regulated by internal cytoskeletal cycles of actin polymerization or blebbing~\citep{yoshida2006dissection}. Models of expanding-contracting particles have been studied recently, focusing on the collective motion of passive particles \citep{tjhung2017discontinuous} or on the single cell behaviour \citep{tarama2018mechanics}.
Our model makes a step further by considering fluctuating volume and active particles, and can be derived as a limiting case of the one described in Ref.~\citep{tarama2018mechanics}, where  
the two elements of the cell are assigned periodically two finite friction coefficients with a generic phase shift.

The active force implemented in Eq.~(\ref{eq:extension}) is not constant, but depends linearly on the length of the cell.
This choice results in a coupling between shape deformations and migration that is found in biological cells~\citep{lauffenburger1996cell}:  circular cells receive a weaker active force, thus move slower, while elongated cells move faster. Since a cell in a dense environment will have a shorter length, the reduction of the active force in Eq.~(\ref{eq:extension}) also models the inhibition of locomotion real cells experience when coming into contact with surrounding cells. This mechanism is often referred to as Contact Inhibition of Locomotion (CIL), and results from a complex interaction that involves both mechanical and chemical signals among neighboring cells~\citep{mayor2010keeping}.
In our model, the details of CIL are enclosed in the shape of the dipole force $\Fmot$ described above.
As discussed in Ref.\citep{schnyder2017collective}, this is sufficient to correctly reproduce the effects of CIL on the collective migration, which has been investigated in several works~\citep{carmona2008contact,abraham2009ve,teddy2004vivo}.

In Ref.~\citep{schnyder2017collective} the model is studied in the limit of $\DT \to 0$ where no cyclic dynamics is performed, the effective friction coefficients are $2 \xi$, and the motility force is applied to the front disk exclusively. Given $\Rmax$ the maximum physical extension of a cell, and $\kappa$ the parameter controlling its stiffness, it is possible to derive in the limit  of $\DT \to 0$ the steady-state cell's length and speed
\begin{equation}
r^0_{\mathrm{ss} }  = \Rmax \sqrt{1 - 2 \kappa / m}, \;\;\;\; v^0_{\mathrm{ss} } = r^0_{\mathrm{ss} }  m / (4 \xi).
\end{equation}
From these a migration time scale can be defined as the time it takes for a cell to travel a distance equal to its maximum length, $\taumig = \Rmax / v^0_{\mathrm{ss} }$. For more information on how these quantities change upon changing $\DT$, see Supplementary Figure S1 and Supplementary Movie 1.

\subsection*{Interactions}

The passive forces in Eq.~(\ref{eq:eq_motion}) are
\begin{align}
\mathbf{F}_i^\mathrm{f} &=
  +\mathbf{F}_{\mathrm{Fene}}(\mathbf{r}_i) 
  + \sum_{j \neq i}  \mathbf{F}^{\mathrm{f} \mathrm{f}}_{\mathrm{WCA}}(\mathbf{r}_{ij}^{\mathrm{f} \mathrm{f} }) + \mathbf{F}^{\mathrm{f} \mathrm{b}}_{\mathrm{WCA}}(\mathbf{r}_{ij}^{\mathrm{f} \mathrm{b} })\\
\mathbf{F}_i^\mathrm{b} &=
  -\mathbf{F}_{\mathrm{Fene}}(\mathbf{r}_i) 
  + \sum_{j \neq i}  \mathbf{F}^{\mathrm{b} \mathrm{f}}_{\mathrm{WCA}}(\mathbf{r}_{ij}^{\mathrm{b} \mathrm{f} }) + \mathbf{F}^{\mathrm{b} \mathrm{b}}_{\mathrm{WCA}}(\mathbf{r}_{ij}^{\mathrm{b} \mathrm{b} }),
\label{eq:passive_forces}
\end{align}
where $\mathbf{F}_{\mathrm{Fene}}$ is the intracellular force between the front and the back disk, $\mathbf{F}^{\alpha \beta}_{\mathrm{WCA}}$ is the interaction between disk $\alpha$ of cell $i$ with disk $\beta$ of cell $j$,  $\mathbf{r}_i$ is the vector from  the back to the front disk of cell $i$, and $\mathbf{r}^{\alpha \beta}_i$ is the vector from disk $\alpha$ of cell $i$ to disk $\beta$ of cell $j$.
The second term on the right hand side of Eq. (\ref{eq:eq_motion}) represents the internal forces within the cell.
The intracellular force is modeled via the Fene force~\citep{warner1972kinetic},
\begin{equation}
  \mathbf{F}_{\mathrm{Fene}}(\mathbf{r}) = \frac{-\kappa}{1 - (r / R_{\mathrm{max}})^2 } \, \mathbf{r},
\end{equation}
where $\kappa$ represents the elasticity of the cell and $R_{\mathrm{max}}$ is its maximum physical extension.
The interaction between disks pertaining to different cells is modeled via the repulsive Weeks-Chandler-Andersen force~\citep{weeks1971role}
\begin{equation}
  \mathbf{F}^{\alpha \beta}_{\mathrm{WCA}}(\mathbf{r})= -24 \epsilon \left[ 2\left( \frac{\sigma_{\alpha \beta}}{r} \right)^{12} + \left( \frac{\sigma_{\alpha \beta}}{r} \right)^6 \right]  \frac{\mathbf{r}}{r^2}, \,\,\,\, r < r_{\mathrm{cut}}
\end{equation}
with $\sigma_{\alpha \beta} = (\sigma_{\alpha} + \sigma_{\beta})/2$, and $r_{\mathrm{cut}} = 2^{1/6} \sigma_{\alpha \beta}$.
The relative size of the front and back disk determines the shape of the cell, which we fix to $\sigma_{\mathrm{b}} / \sigma_{\mathrm{f}} = 0.8$.

\subsection*{Simulations}

We integrate Eq.~(\ref{eq:eq_motion}) employing the Euler-Maruyama integrator, with timestep $4 \cdot 10^{-3}$ in Lennard-Jones units, corresponding to approximately $3.75\cdot 10^{-4} \tau$ (with $\tau$ is the migration time scale of the model at $\DT = 0$). Cells are initialized by positioning the front disk on a cubic lattice, and assigning random positions to the back disk around the front disk, taking care of not overlapping different cells. For the simulations with finite $\DT > 0$, the initial internal time $t_i$ and stage $s_i$ are assigned randomly using the pseudo-random number generator of the GNU Scientific Library (GSL), so that cells' cycles are not synchronized. We use a square box for each simulation, and implemented periodic boundary conditions.

For each simulation, data is collected only after the steady state is reached. Each point in the figures shown is obtained by averaging over 15 independent simulations, excluding those simulations that displayed an exceptionally long time to reach the steady state, because of the formation of long lived metastable vortices, see Supplementary Movie 5. 

\subsection*{Correlation functions}

In order to quantify local correlation/anticorrelation of the internal time $t_i$ and stage $s_i$ of a test cell $i$ with its surrounding cells, we compute the respective spatial correlation functions $C(x',y')$ and $S(x',y')$ as follows
 \begin{equation}
\begin{aligned}
C(x', y') = \Bigl<  \sum_{i = 1}^{\Ncell} \sum_{j = i + 1}^{\Ncell} & \cos\left[\frac{\pi}{\DT}( t_{ij} + \DT s_{ij}) \right] \times{} \\
  & \times \delta( x'_{ij} - x') \delta( y'_{ij} - y')  \Bigr>
\end{aligned}
\label{eq:correlation_stagetime}
\end{equation}
and
\begin{equation}
\begin{aligned}
 S(x', y') = \Bigl<  \sum_{i = 1}^{\Ncell} \sum_{j = i + 1}^{\Ncell} & \; \left[2 \left( s_i + s_j \, \mod  2 \right) - 1 \right] \times \\ 
  & \times \delta( x'_{ij} - x') \delta( y'_{ij} - y')  \Bigr>
\end{aligned}
\label{eq:correlation_stage}
\end{equation}
where $z_{ij} = z_j - z_i$, and $x'_j, y'_j$ are the spatial coordinates of the front disk of cell $j$ in the reference frame of cell $i$, in which the $\mathbf{\hat{y}'}$ axis coincides with the cell's orientation, centered at the front disk.
Each cosine in the summation of $C(x',y')$ can give values up to $+1$ if the cell pair $(i, j)$ has close to no difference in the internal times or a difference close to the period $\DT$ (correlated), while giving values close to $-1$ if the difference is approximately $\DT/2$ (anticorrelated).
From this calculation alone, it is not known whether two cells with (anti)correlated internal times possess the same or the opposite stage: $S(x', y')$ is useful to distinguish between these two different cases.

\section*{Results}

\subsection*{Collective migration without crawling}
\begin{figure*}[t!]
  \centering
  \includegraphics{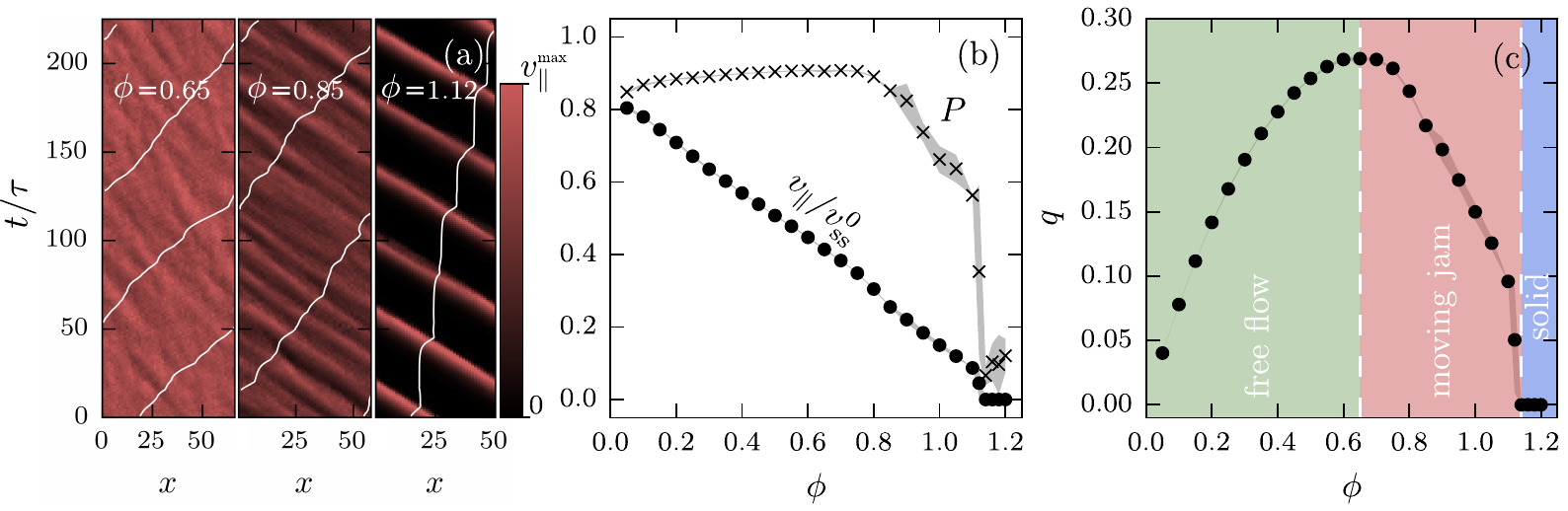}
  \caption{(a) Kymograph of the migration speed $v_{\parallel}(x)$ for packing fractions $\phi = \{ 0.65,\, 0.85,\, 1.12\}$. The migration direction is along the $y$ axis. The emergence of speed waves is indicated by the peak of $v_{\parallel}(x)$ moving towards negative $x$ in time, forming linear bands in the kymograph. The white line represents the trajectory of a single cell, which is moving towards larger $x$ in time, thus in the opposite direction of the velocity waves. This is a typical feature of velocity waves in a traffic jam. (b) Discs represent the parallel speed $v_{\parallel}$ (CIL) as a function of packing fraction $\phi$, rescaled with the steady state speed  $v^0_{\mathrm{ss}}$, and crosses represent the polar order parameter $P$. The shaded area indicates the standard deviation of the data points. (c) Flux of cells exhibiting CIL shows a maximum at the critical packing fraction $\phi_c  \approx 0.65$, where the system transitions from the free flow regime to the moving jam regime, and solidifies beyond $\phi_{\mathrm{s}} \approx 1.14$. The model without CIL exhibits no net flux since there is no collective alignment. See also Supplementary Movie 2 and 3.}
  \label{fig:nocycles_traffic}
\end{figure*}

We now focus on the emergence of aligned migration and the formation of velocity waves.
We first consider the model for $\DT = 0$ (no crawling) and show that the collective migration shows similar behavior to that of traffic flow. In the next section, using the model at $\DT = 0$ as a reference, we study how the introduction of an explicit crawling dynamics through the use of a finite $\DT>0$ affects the collective behavior of the cellular tissue.

In order to characterize the collective alignment, we compute the polar order parameter,
\begin{equation}
  P = \langle |\mathbf{e}| \rangle, \qquad 
  \mathbf{e} = \frac{1}{\Ncell}\sum_{i = 1}^{\Ncell} \frac{\rbfi}{\left| \rbfi \right|},
\label{eq:migration_e}
\end{equation}
where $\Ncell$ is the total number of cells, and $\langle\dots\rangle$ denotes an ensemble average performed by averaging over time and independent runs, each starting from different and randomized initial conditions, see \emph{Methods}. 
$P$ can take values between zero (random) and unity (perfect alignment). For non-negligible values of $P$, we also compute the average speed along the instantaneous migration direction $\mathbf{e}$,
\begin{equation}
  v_{\parallel} =  \left< \frac{1}{\Ncell} \sum_{i = 1}^{\Ncell} \mathbf{v}_i  \cdot 
    \frac{\mathbf{e}}{\left| \mathbf{e} \right|} \right>,
  \label{eq:vparallel}
\end{equation}
where $\mathbf{v}_i = (\mathbf{v}^{f}_i + \mathbf{v}^{b}_i) / 2$ is the center of mass velocity of cell $i$.

We compute $P$ and $v_{\parallel}$ for a range of densities.
In order to have a dimensionless quantity expressing the density, we use the packing fraction $\phi = N A_{\mathrm{ss}} / L^2$, where $A_{\mathrm{ss}}$ is the area of a single cell in the steady state,  and $L$ is the linear length of the box. Using this definition, freezing occurs at $\phi \approx 1.15$ in the model with $\DT = 0$.
At high packing fractions ($\phi \geq 0.65$) the flow of the tissue develops velocity waves which resemble the waves observed during traffic congestion, where the wave vector is anti-parallel to the migration direction, see Supplementary Movies 2 and 3. Velocity waves have been observed in experiments of tissue expansion~\citep{serra2012mechanical} and in the aggregation of Dictostelium discoideum during starvation~\citep{alcantara1974signal}. In the latter, velocity waves are accompanied by waves of chemoattractants that are emitted by the cells during starvation.

In our model, velocity waves are purely related to cell-cell interaction and shape deformation upon collision.
More generally, the formation of traveling bands in systems of active particles has been observed for Vicksek-like particles~\citep{chate2008modeling}, soft deformable active particles~\citep{yamanaka2014formation}, and colloidal rollers~\citep{bricard2013emergence}, among others. In order to visualize the waves, a convenient plot is the kymograph of the velocities [see Fig.~\ref{fig:nocycles_traffic}(a)]. The kymograph is constructed by averaging the cell's velocities along the $y$ direction, orthogonal to the progagation direction $x$ of the waves, for different times. As $\phi$ is increased a structure of bands emerges, where layers of slow particles alternate with layers of faster particles.
The slope of the bands in the $t$-$x$ plot indicates the direction of the wave, which is opposite to the direction of individual cells, as indicated by the white line in Fig.~\ref{fig:nocycles_traffic}. It is instructive to note that aligned migration and velocity waves arise only when using the motility force that models CIL. The qualitative reason is that a strong dependence of the local velocity on the local density is present, cf. Fig.~\ref{fig:nocycles_traffic}(b). This is critical for the emergence of traffic waves, which arise from a positive feedback between density and velocity~\citep{kerner1993cluster}: if the speed decreases locally, the density will increase locally, which in turn will cause the speed to lower further, and so on.

Clearly, the formation of traffic waves indicates congestion, which is not beneficial for migration.
In order to characterize the efficiency quantitatively, we compute the adimensional flux $q$ of cells
\begin{equation}
q =  v_{\parallel} \phi / v^0_{\mathrm{ss}}.
\end{equation}
In Fig.~\ref{fig:nocycles_traffic}(c) the flux $q$ is plotted against area fraction $\phi$. The observed behavior is analogous to the one obtained from studies of car traffic~\citep{kerner1999physics}, where two regimes can be distinguished by the sign of $\partial{q}/\partial{\phi}$: the \emph{free flow} and the \emph{moving jam}.
In the free flow regime, the flux increases as the density increases and no congestion arises. The flux $q$ reaches a maximum value at a critical packing fraction $\phi_c \approx 0.65$, after which velocity waves start to emerge. 
By increasing the density beyond $\phi_c$, cells start to cluster into waiting lines and the flux is reduced (\emph{moving jam} regime). The shape of the flux-versus-density curve, usually referred to as the fundamental diagram in traffic flow studies, acquires a parabolic shape. This behavior is typical for transport processes dominated by excluded volume and can be understood within the Totally Asymmetric Exclusion Process (TASEP)~\citep{spitzer1970interaction}, with the difference that here the cells are not confined to one dimension.

\begin{figure*}[t!]
  \centering
  \includegraphics{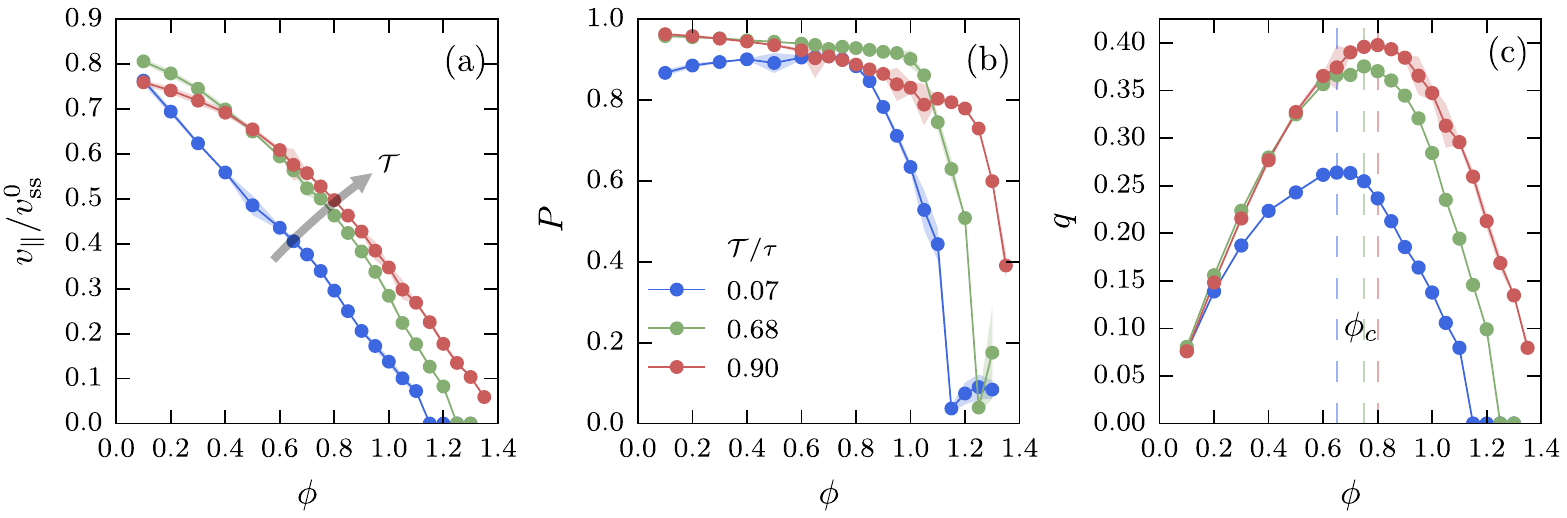}
  \caption{(a) Average speed along the direction of migration $v_{\parallel}$ rescaled with $v^0_{\mathrm{ss}}$, plotted against packing fraction $\phi$ for $\DT = \{ 0.07, 0.68, 0.90\} \taumig$. Introducing an expansion-contration dynamics induces a non linear trend of $v_{\parallel}$ for which an increase in $\phi$ is accomodated with a softer decrease in $v_{\parallel}$, signaling some kind of emergent cooperativity in the system. (b) Polar order paramenter $P$ for the system  (see eq.~\ref{eq:migration_e}): as $\DT$ is increased, the system is able to maintain collective alignment for increasingly higher densities. (c) The flux of cells is significantly enhanced when increasing the duration of the cyclic dynamics, and the critical packing fraction $\phi_c$ shifts from $0.65$ to $0.8$. See also Supplementary Movie 4.}
  \label{fig:cycles_optimal}
\end{figure*}

\subsection*{Collective migration with crawling}

We now turn our attention to the model where the cyclic dynamics of extension-contraction are explicitly included.
By switching $\DT$ to finite non-zero values, the dependence of the average migration velocity $v_{\parallel}$ on $\phi$ changes from linear to curved [Fig.~\ref{fig:cycles_optimal}(a)]. This change is a consequence of the system being able to accommodate more cells given the same space, and with a smaller decrease in migration speed, which implies emerging cooperativity. Velocity waves are now absent and the system is able to maintain high alignment for increasingly higher packing fractions (see Fig.~\ref{fig:cycles_optimal}(b) and Supplementary Movie 4). The resulting flux of cells is enhanced [Fig.~\ref{fig:cycles_optimal}(c)], with $\phi_c$ shifted to a higher value of $\phi_c\simeq 0.80$. At high packing fractions it is even possible to transition from the zero-flux solid to a traveling tissue with finite flux, see Supplementary Figure S2. Note that a similar behavior has been reported in lattice gas models of traffic flow, where the introduction of a stepping kinetic cycle to the TASEP has been found to enhance the flux of molecular motors~\citep{klumpp2008effects,ciandrini2014stepping}. In those systems, the time spent queuing is used for advancing the internal state, so that molecular motors are immediately ready to move once free space is available, thus enhancing the flux. Although the effect is similar, the microscopic origin in our system is quite different, as we are going to show at the end of this section.

What, then, is the origin of this collective migration in the absence of any explicit communication mechanism?
We observe that for finite $\DT$, cells tend to spontaneously arrange themselves in such a way that their neighboring cells are either synchronized or anti-synchronized in their expansion/contraction stage, depending on their relative spatial location. This is surprising since the internal clocks of cells run independently. Such spontaneous synchronization facilitates a cell's movement within its local neighborhood. In order to quantify this phenomenon, we compute two spatial correlation functions: of the internal time $t_i + s_i \DT/2$ and of the internal stage $s_i$, both as a function of cell separation (denoted $C$ and $S$, respectively, for details see \emph{Methods}). We stress that these quantities evolve independently in the simulation since there is no explicit interaction between $t_i$ or $s_i$ of different cells.

\begin{figure*}[t!]
  \centering
  \includegraphics[width=1.0\textwidth]{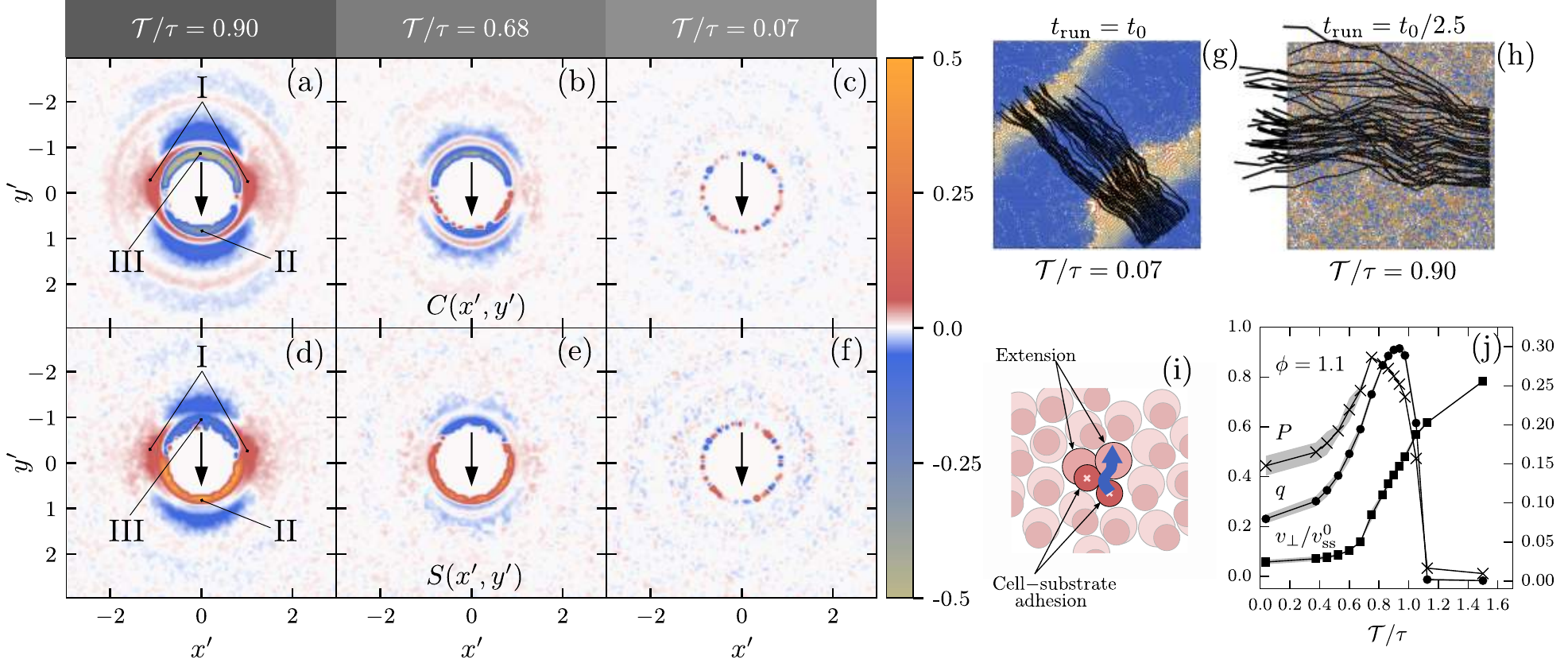}
  \caption{(a-c) Correlations $C(x', y')$ and (d-f) $S(x',y')$ for three different cycle durations, showing the emergence of the correlation/anti-correlation pattern in space, responsible for the enhancement of the flux. The black arrows indicate the direction of the migrating cell. (g,h) Trajectories of a subset of cells (black lines) at $\phi = 1.1$ for $\DT / \taumig = 0.07$ and $0.90$ respectively, plotted on top of representative configurations of the system. The cells are colored according to their lenght $\left| \mathbf{r}^{\mathrm{bf}}_i \right|$ from blue (shortest) to red/white (longest). $t_0$ indicates the time duration for producing the trajectories in (g), and $t_0 / 2.5$ for (h). (i) The mechanism of "lane change", which is facilitated by the adhesions to the substrate during crawling. (j) Flux of cells $q$, $v_{\perp}$ and polarity $P$ against cycle length $\DT / \taumig$ for a fixed packing fraction $\phi = 1.1$. The system reaches its maximum flux for $\DT \approx 0.90 \taumig$ while keeping a high collective alignment, and increasing $v_{\perp}$. By further increasing the cycle length, the system loses alignment and no net flux is observed at $\DT / \taumig \approx 1.1$. See also Supplementary Movie 4.}
  \label{fig:local_synchronization}
\end{figure*}

The spatial coordinates $x'$ and $y'$ are relative to the frame of reference of a tagged cell, where the $\mathbf{\hat{y}'}$ axis coincides with its orientation, and the center is placed at its front disk.
Fig.~\ref{fig:local_synchronization}(a-f) presents the resulting correlations $C(x',y')$ and $S(x',y')$ for packing fraction $\phi = 1.1$ and for three different cycle durations $\DT =  \{0.07, 0.68, 0.90\} \taumig$.
Blue and green indicate anti-correlation, red and orange indicate correlation, and white indicates no correlation.
At $\DT \ll \taumig$ both $C(x', y')$ and $S(x',y')$ show no particular structure except for concentric circles due to the packing of cells. As we increase $\DT$ to values close to $\taumig$, a pattern develops at short distances from the cell.
First, we note that neighboring cells migrating side by side (regions I in Fig.~\ref{fig:local_synchronization}(a,d) ) are more likely to be in the same stage with similar internal times.
More interestingly, cells directly in front (regions II) are found to be mostly anti-correlated with respect to the internal time, but correlated with respect to stage. Hence, if cell $i$ is at the start (end) of stage $s_i$, a cell $j$ directly behind will most likely be at the same stage $s_j = s_i$ but at the end (start) of it.
Cells directly behind the tagged cell (regions III) are anti-correlated both in the internal time and in the stage, meaning that if cell $i$ is found expanding (contracting) then cell $j$ directly in front will most likely be contracting (expanding). One could expect the plots of $C(x',y')$ and $S(x',y')$ to be symmetric both in the $y'$ and $x'$ axis, since when implementing the calculation one treats together the cell pairs $(i,j)$ and $(j,i)$ and $\mathbf{r}_{ij} = -\mathbf{r}_{ji}$. However, this is not the case since one has to first rotate the coordinates to the reference frame ($\mathbf{r}_{ij} \to \mathbf{r}'_{ij}$) of each cell.

It is worth noting that, although cells in the system with $\DT \ll \taumig$ keep their neighbors for very long times (since cells spend a long time queuing), the correlation pattern is zero in constrast to the case $\DT = 0.9 \taumig$. 
In the latter, cells are dynamically changing neighbors all the time, but show remarkably robust ability to maintain the pattern of correlation we observe in Fig.~\ref{fig:local_synchronization}. Fig.~\ref{fig:local_synchronization} shows also that a weaker, but nonetheless non-zero, correlation/anti-correlation pattern is present beyond the first coordination shell. Although the structure we observe is short-ranged and does not extend beyond about twice the cell diameter, it is sufficient to globally enhance the migration of the tissue.

What is the microscopic explanation for the enhancement of collective migration and for the spontaneous synchronization of cells, linking the two observations together? First, it is instructive to notice that trajectories of single particles differ qualitatively when increasing $\DT$ to $0.9 \taumig$: instead of queuing and then moving in straight lines when the velocity waves arrive, cells tend to move in a zig-zag fashion, as shown in Fig.~\ref{fig:local_synchronization}(g,h).
The reason for the zig-zag trajectory is sketched in Fig.~\ref{fig:local_synchronization}(i) and shown in Supplementary Movie 4: if a cell during extension finds another extending cell in front, volume exclusion with the back disk allows the first cell to slip past the second, provided $\DT$ is large enough and enough space is available.
Such a mechanism is reminiscent of lane changes in the traffic flow of cars, and results in a higher probability to find synchronized cells crawling side by side, and asynchronized cells in the front or back.
We stress that the adhesion to the substrate is crucial for this lane changing mechanism, since during the crawling cycle cells have their back and front disk's position fixed sequentially.

We can quantify the effect by computing the cell velocity along the direction orthogonal to the collective migration direction
\begin{equation}
v_{\perp} =   \left< \frac{1}{\Ncell} \sum_{i = 1}^{\Ncell} \left| \mathbf{v}_i  - \mathbf{v}_i \cdot \frac{\mathbf{e}}{{\left| \mathbf{e} \right|}} \right|  \right>
\label{eq:vperp}
\end{equation}
Fig.~\ref{fig:local_synchronization}(j) shows the results for $v_{\perp}$, the flux $q$, and polarity $P$ for a fixed packing fraction $\phi = 1.1$ as a function of $\DT$. As expected, $v_{\perp}$ shows an increase in agreement with the increase of flux $q$, which becomes maximal when $\DT \approx 0.9 \taumig$ reaching a value $55\%$ higher than in the case $\DT = 0$. The plot shows also a steep decay of the flux $q$ and polarity $P$ immediately after $\DT \approx \taumig$, where the tissue ceases to exhibit coherent motion. In the latter case, the internal cycle of the cell is too long and alignment is disrupted.


\section*{Discussion}

In this work, we have shown how the introduction of an internal cycle for the movement of single cells strongly influences the collective migration properties of the whole tissue. The internal cycle we use models the ameboid migratory phenotype, which is typical for a wide class of cells and consists of a discrete series of conformational changes of the cell that result in its directed motion. We find that the flux of cells is enhanced and optimized by increasing the duration of the cycle to match a specific value of around $0.9 \taumig$, where $\taumig$ is the migration timescale of the cell in the limit of very short cycles. We have shown that this behavior can be clarified microscopically by the increased ability of cells to perform ``lane changes'' while crawling, which in turns leads to the emergence of a local correlation/anti-correlation pattern of the cycles of expansion and contraction. The result is an emergent global cooperation among cells in the tissue, through which they achieve better migration properties, which manifests itself through a larger flux without the traffic waves indicating congestion.

Our insights can potentially be exploited to influence and steer cell migration. Through tuning the duration of the internal cycle of cells it becomes possible to either enhance the migration of cells, or, to the contrary, diminish the migration of non-desirable (malignant) cells. Influencing the cycle duration might be accomplished through manipulating the substrate, \emph{e.g.}, periodic stretching of the substrate is able to reorient solitary crawling cells~\citep{Janmey2007,Jungbauer2008,Iwadate2013,Livne2014a,Chen2015a,Okimura2016,molina2018mechanosensitivity}. A first step will be to verify our results in experiments that track cells individually and discriminate between expanding and contracting cells, which allows to compute the correlations $C(x',y')$ and $S(x',y')$ introduced here. Such investigations will yield insights into the importance of our purely physical mechanisms as compared to other well-known mechanisms like chemotaxis, cell division, and cell-cell adhesion. 

\section*{Supporting Material}

An online supplement to this article can be found by visiting BJ online at \emph{link}.

\section*{Author contributions}

MC, TS, and RY designed research. MC, SKS, and JJM worked on the model and simulations, MC performed the simulations and analyzed the data. All authors contributed to interpreting the results and writing the paper.

\section*{Acknowledgments}

MC is funded by the DFG through the Graduate School ``Materials Science in Mainz'' (GSC 266) and the collaborative research center TRR 146.  MC gratefully acknowledges GSC 266 for funding the research stay in Kyoto during which part of this work was carried out. RY acknowledges the Japan Society for the Promotion of Science (JSPS) KAKENHI (17H01083) grant, and the JSPS bilateral joint research projects. JJM acknowledges the Japan Society for the Promotion of Science (JSPS) Wakate B (17K17825) grant. The calculations were performed using the computational facilities of the University of Kyoto and the supercomputer Mogon at the Johannes Gutenberg University Mainz.



\newpage
\appendix

\section{Single cell's cyclic motion}

\begin{figure*}[h]
  \centering
  \includegraphics{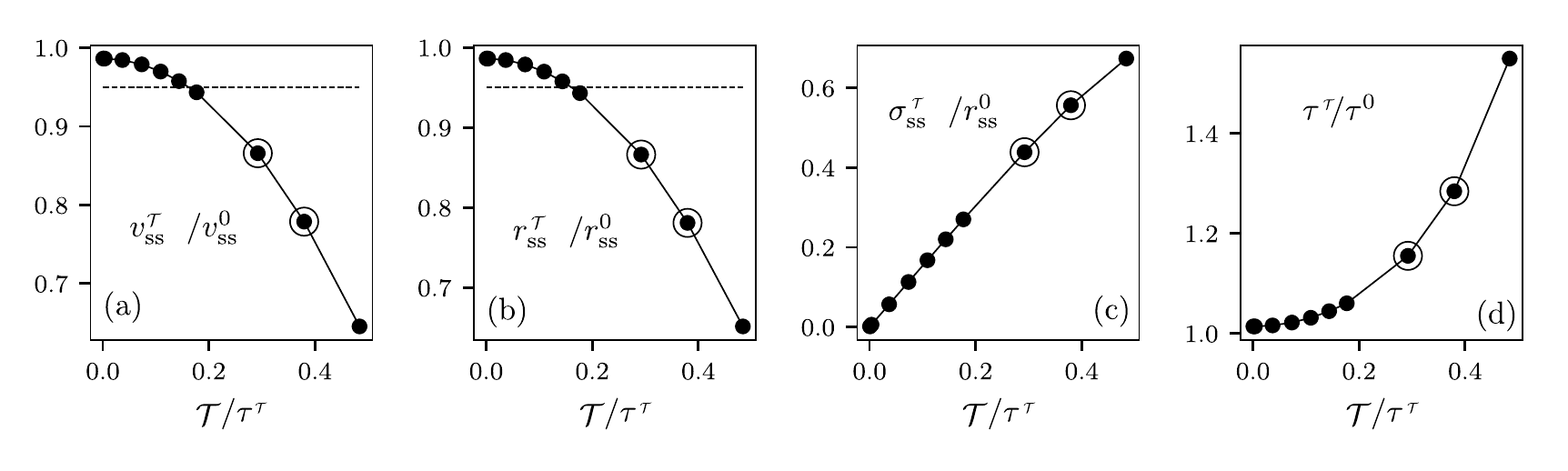}
  \caption{Motion of the single cell under periodic extension-contraction cycles ($\DT > 0$). Quantities are normalized to the model with $\DT = 0$, where no explicit cycle is performed. (a) Steady state speed $v^{\DT}_{\mathrm{ss}}$, averaged over one cycle. (b) steady state length of the cell $r^{\DT}_{\mathrm{ss}}$, averaged over one cycle. (c) standard deviation of the cell's length $\sigma^{\DT}_{\mathrm{ss}}$ during a cycle. (d) migration time $\tau^{\DT}$. Full lines are a guide to the eye, and the dashed lines in (a,b) indicate where the values drop below $95\%$ compared to the case $\DT = 0$.}
  \label{fig:single_cell_migration}
\end{figure*}

By changing the duration of the contraction-extension cycle, the steady state properties of single cell migration change, see Fig~\ref{fig:single_cell_migration} and Supplementary Movie 6.
As mentioned in the main text, the steady state velocity of the cell $v^0_{\mathrm{ss}}$ decreases, while the collective migration counterintuitively improves, see Main Text and Supplementary Movies 3 and 4.
For cycle duration $\DT \approx 0.9 \taumig$, which is where the collective migration is enhanced the most at $\phi = 1.1$, $v^0_{\mathrm{ss}}$ is $20\%$ lower compared to the case $\DT = 0$.
The averaged (over one cycle) length of a cell at steady state, $r^0_{\mathrm{ss}}$, is also decreased upong incresing $\DT$.
While the average length of the cell is shorter, the amplitude of oscillation is higher.
Since the maximum length of the cell $\Rmax$ is fixed and the steady state speed $v^0_{\mathrm{ss}}$ increases, the migration time $\taumig = \Rmax / v^0_{\mathrm{ss}}$ increases as well.

\section{Transition to a moving tissue}

\begin{figure}[t]
  \centering
  \includegraphics{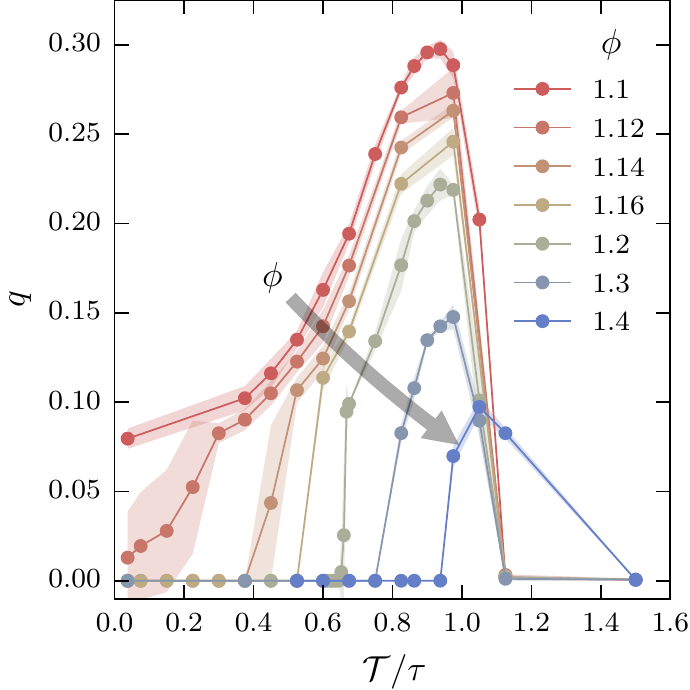}
  \caption{Flux $q$ of the tissue of cells for several packing fractions $\phi$, according to the colors, for a range of cycle durations $\DT$}
  \label{fig:transition}
\end{figure}

As shown in the main text in Fig.~4(j) for $\phi = 1.1$, increasing $\DT$ leads to better migration with enhanced flux.
We show here that not only the flux $q$ can be increased, but for packing fractions at which the system at $\DT = 0$ is in the solid phase, $\DT$ can drive a phase transition from the static solid into a migrating tissue.
In Fig.~\ref{fig:transition}, we show the flux $q$ as function of $\DT$ for increasing packing fractions.
For all packing fraction studied above  $\phi = 1.1$, we observe a sharp transition from $q = 0$ to a finite value $q>0$.
The data for $\phi = 1.2$ (represented in green) demonstrates how steep the transition is, even when increasing $\DT$ of just 10 simulation time steps.
This finding shows how conformational changes in cell tissues at high density are sufficient to trigger collective migration.

\section{Metastable vortices}

In systems with CIL it is possible to observe long lasting metastable vortices before the occurrence of the steady state collective migration.
Such states are characterized by the presence of two vortices of opposite charge for the system sizes we have so far considered (up to $\Ncell = 10^4$),  see Supplementary Movie 5. The occurrence of vortices depends solely on the initial conditions, since the model does not employ any random noise, and it is observed more frequently (from 10\% to 50\% of the samples) in the region of packing fractions $0.65 \leq \phi \leq 1.0$. For this reason, we have excluded from the presented data those runs where the vortices where particularly stable to last for the majority of the simulation time.

\end{document}